\documentclass[aps,onecolumn,11pt,floatfix,altaffilletter,
               tightenlines,showpacs,showkeys,notitlepage,nofootinbib]{revtex4-1}

\usepackage[colorlinks=true,citecolor=blue,linkcolor=blue]{hyperref}
\usepackage[normalem]{ulem}
\usepackage{amsmath,amssymb}
\usepackage{epsfig}
\usepackage{graphicx}               
\usepackage{url}
\usepackage{color}
\usepackage{multirow}
\usepackage{placeins}
\usepackage[dvipsnames]{xcolor}
\usepackage{epstopdf}

\usepackage{tikz}
\usetikzlibrary{trees}
\usetikzlibrary{decorations.pathmorphing}
\usetikzlibrary{decorations.markings}

\definecolor{jblue}  {RGB}{20,50,100}
\definecolor{npurple}  {RGB} {153, 51, 204}
\definecolor{wred}   {RGB}{217,0,56}
\definecolor{white}   {RGB}{255,255,255}

\definecolor{korange}   {RGB}{235, 80,  43}
\definecolor{korange2}   {RGB}{245, 100,  63}
\definecolor{kyelloworange}   {RGB}{255, 210,  110}
\definecolor{kyelloworange2}   {RGB}{240, 170,  90}
\definecolor{kred}   {RGB}{204,  102, 153}
\definecolor{kpurple}   {RGB}{153,  61, 190}
\definecolor{kpurplelight}   {RGB}{213,  161, 230}

\usepackage[capitalize]{cleveref}
\usepackage[caption=false]{subfig}
\usepackage{lipsum}
\definecolor{red}{rgb}{1.0, 0, 0}
 \usepackage{gensymb}
 
\allowdisplaybreaks

\bibpunct{[}{]}{,}{n}{}{,}

\newcommand{\GeV}{\,\mathrm{GeV}}
\newcommand{\TeV}{\,\mathrm{TeV}}
\newcommand{\PeV}{\,\mathrm{PeV}}
\newcommand{\EeV}{\,\mathrm{EeV}}
\newcommand{\pcs}{\,\mathrm{parsecs}}
\newcommand{\pc}{\,\mathrm{parsec}}

\newcommand{\Mpc}{\,\mathrm{Mpc}}
\newcommand{\Gpc}{\,\mathrm{Gpc}}


\begin{document}

\title{IceCube Flavor Ratios with Identified Astrophysical Sources: \\
Towards Improving New Physics Testability}

\author{Vedran Brdar}   \email{vbrdar@mpi-hd.mpg.de}
\author{Rasmus S. L. Hansen}   \email{rasmus@mpi-hd.mpg.de}
\affiliation{Max-Planck-Institut f\"ur Kernphysik,
       69117~Heidelberg, Germany}

\begin{abstract}
\noindent
Motivated by the discovery of the first high-energy astrophysical neutrino source, the blazar TXS 0506+056, we revisit the IceCube flavor ratio analysis. Assuming large statistics from identified blazars, collected in the forthcoming years by the IceCube detector and its successor IceCube-Gen2, we demonstrate that the constraints on several new physics scenarios in which the baseline dependent terms in neutrino oscillation probabilities are not averaged, can be improved. As a representative case, we consider pseudo-Dirac neutrinos while neutrino decay is also discussed. 
\end{abstract}

\maketitle

\section{Introduction}
\label{sec:intro}
\noindent
Multi-messenger astronomy has recently become one of the most exciting fields in astroparticle physics. The recent discovery of gravitational waves~\cite{Abbott:2016blz} and electromagnetic radiation from a neutron star merger~\cite{GBM:2017lvd}, as well as the detection of the first high-energy neutrino with the accompanying $\gamma$-rays 
from blazar TXS 0506+056~\cite{IceCube:2018dnn} have indicated a great potential for 
an improved understanding of various astrophysical sources as well as hinted at the exciting possibility for more similar observations in the near future. The discovery of the first high-energy ($E_\nu>100 \TeV$) neutrino from the blazar TXS 0506+056 is of great importance for neutrino physics since the previously identified astrophysical neutrino sources, Supernova 1987A~\cite{Hirata:1987hu} and the Sun, only have led to the detection of $E_\nu\sim$ $1$-$10$ MeV  neutrinos. Moreover, it is now estimated at $>3\sigma$ that TXS 0506+056 has previously emitted several more high energy neutrinos. Despite current difficulties with explaining these previously detected events with blazar models, more events, not only from TXS 0506+056 but other blazars as well, are expected to be detected by IceCube in the future. With a planned upgrade to IceCube-Gen2, the accumulation of events would be around ten times faster than for the present observatory, and neutrinos from known blazar sources could eventually be analyzed using techniques previously only applied to the full samples of high-energy starting events (HESE) and through-going muons (TGM).

One such analysis is the study of flavor composition for high-energy neutrinos arriving at Earth. This topic has been widely studied~\cite{Barenboim:2003jm, Cirelli:2004cz, Xing:2006uk, Lipari:2007su, Pakvasa:2007dc,        Blennow:2009rp, Esmaili:2009dz, Lai:2009ke, Choubey:2009jq, Bhattacharya:2010xj, Hollander:2013im, Rajpoot:2013dha, Mena:2014sja, Xu:2014via, Fu:2014isa, Palomares-Ruiz:2015mka, Palladino:2015zua, Aartsen:2015ivb, Kawanaka:2015qza, Palladino:2015vna, Arguelles:2015dca, Bustamante:2015waa, Aartsen:2015ita, Brdar:2016thq, Ahlers:2018yom,Bazo:2009en,Shoemaker:2015qul,deSalas:2016svi,Denton:2018aml}, especially after the discovery of the first high-energy neutrinos by the IceCube collaboration in 2013. In this paper we focus our interest on the neutrino flavor composition from individual blazar sources in the presence of new physics affecting neutrino propagation. More specifically, we explore scenarios in which the source distance enters in the calculation of neutrino flavor conversion.
For instance, in the pseudo-Dirac neutrino model \cite{Kobayashi:2000md, Beacom:2003eu, Esmaili:2012ac}, the operative mass squared difference $\Delta m^2$ between active and sterile neutrinos can be tiny, and the effects from oscillations between such states could be observable in neutrino telescopes.
This is in contrast to complementary new physics models\footnote{e.g. sterile neutrino with $\Delta m^2 \sim \mathcal{O}(1)\,\text{eV}^2$ \cite{Brdar:2016thq} or non-standard interactions where the oscillation phase depends on the standard mass squared differences $\Delta m^2 \sim 10^{-3}\,,10^{-5}\, \text{eV}^2$ \cite{Arguelles:2015dca}.} in which the terms dependent on propagation distance get averaged because the baseline exceeds the coherence length.

In the absence of decoherence effects, not knowing the baselines limits the flavor compositions at Earth that are predicted by new physics models.
For models with baseline-dependent oscillation probabilities, it is intuitively clear that the loss of baseline information must impact the flavor composition qualitatively.
Generally, any inferred constraints on new physics parameters from neutrinos with unknown sources are less powerful and robust when compared to scenarios with known sources.

Our working hypothesis is that, in the future, the detected number of high-energy neutrino events from \emph{identified} blazars will be sufficiently large to conduct a flavor ratio analysis. This is feasible because blazars are not transient sources and repeated flaring episodes will lead to a gradual accumulation of high-energy neutrinos.

This paper is organized as follows. In \cref{sec:blazar_astro} we discuss the properties of high-energy neutrino emission from blazars, e.g. the energy spectra and the flavor compositions at the source. In \cref{sec:flavor_ratio} we conduct the flavor composition analysis and demonstrate for the case of pseudo-Dirac neutrinos how knowing the neutrino sources improves the constraints on new physics parameters. Finally, we conclude in \cref{sec:summary}.

\section{Astrophysical High-energy neutrino emission from Blazars}
\label{sec:blazar_astro}
\noindent
A blazar is an active galactic nucleus (AGN) with a relativistic jet pointed in a direction close to the solar system resulting in very high luminosity and large variations in the signal over days, weeks and months. The jet is powered by the accretion of matter onto a supermassive black hole (SMBH). 
Due to beaming of the jet, the luminosity observed on Earth is much larger than the luminosity emitted in the frame of the jet. A consequence of the beaming is that AGNs with a jet pointed towards us look very different from AGNs where the jet is pointed away from us. 
Using the total luminosity of the blazar, the spectral energy distributions (SEDs) in electro-magnetic radiation can be described by the ``blazar sequence'' classification scheme. From low to high luminosity, the classes are: high-frequency BL Lacs, low-frequency BL Lacs, and flat spectrum radio quasars (FSRQs)~\cite{Fossati:1998zn,Ghisellini:1998it}.
In this paper we will focus on BL Lacs which is the class that TXS 0506+056 falls into.

The SED of a blazar has a characteristic two-bump structure with one peak at low energies (UV to X-rays) and a second at high energies (X-rays to $\gamma$-rays)~\cite{Fossati:1998zn}. The low-energy emission is well explained by synchrotron emissions from relativistic electrons in the jet, but there are various mechanisms for producing the high-energy radiation which generally fall into two types of models: leptonic and hadronic. In leptonic models, the high-energy radiation is emitted by  Compton scattering of electrons on photons. The photons are created either by other parts of the blazar or by the electrons themselves through synchrotron radiation. In hadronic models, protons are assumed to be accelerated to high energies. The high-energy radiation is then produced by synchrotron radiation from the proton as well as from the decay products of pions that come from scattering either on photons (photohadron/photomeson) or on nuclei in gas clouds that cross the jet (hadronuclear)~\cite{Boettcher:2013wxa}. Recently, lepto-hadronic models where both mechanisms are at play have become popular~\cite{Halzen:1997hw, Padovani:2015mba, Gao:2016uld, Palladino:2018lov, Murase:2018iyl, Gokus:2018lgx, Sahakyan:2018voh}, especially after the association of a high energy neutrino with TXS 0506+056~\cite{IceCube:2018dnn}.

The production of high energy pions implies that blazars are sources of high energy neutrinos. As a consequence, the observation of neutrinos can be used to distinguish between leptonic and hadronic models~\cite{Gao:2016uld, Murase:2018iyl, Gokus:2018lgx, Sahakyan:2018voh}. 
The neutrino flavor ratio from a blazar is expected to be $(\nu_e:\nu_\mu:\nu_\tau) = (1:2:0)$ both for photomeson production (see e.g. Fig. 12 in Ref. \cite{Hummer:2010vx}) and for hadronuclear production~\cite{Kelner:2006tc} under the reasonable assumption that muons are not losing a significant amount of energy before they decay~\cite{Hummer:2010ai,Kashti:2005qa}. In photomeson models, the density of gas is low, and the main pion production stems from protons interacting with photons. In hadronuclear models, protons interact with a dense gas, but the density is typically assumed to be low enough to allow muons to decay~\cite{He:2018snd}.

The physical extent of the emission regions in blazar jets can be estimated through the typical time scale of variability in the signal of $\sim 1$ day~\cite{Halzen:1997hw}. The variability arises due shock waves going through clumps of matter in the jet with bulk Lorentz factor $\gamma \sim 10$. The clumps must be of the size $\Gamma t c \sim 10^{-2}\pcs$, where the Doppler factor $\Gamma$ transforms between the boosted frame and the observer frame and $\Gamma \approx \gamma$ for observing angles close to the jet direction. The protons are typically Fermi accelerated: the proton crosses the shock front repeatedly and gains energy each time. The maximal proton energy is therefore limited by the size of the accelerating region. With a size of $\sim 10^{-2}\pcs$, the maximal energy is $\sim 10 \EeV$~\cite{Halzen:1997hw}, comparable to the highest energy observed for cosmic rays. 

The neutrino producing region requires the presence of high energy protons and a high density of photons in the case of photomeson production or a dense gas cloud in the case of hadronuclear production. 
For the photomeson process, the appropriate environment is only found close to the SMBH in the most popular models (in other models, neutrino emission occur in layers of the jet further away from the SMBH, see e.g.~\cite{Tavecchio:2014iza,Tavecchio:2014eia,Murase:2014foa}). For the hadronuclear process, the proximity to the SMBH is not as obvious, but the size of the $\gamma$-ray emitting clouds can be estimated through the total duration of a flaring period which is of order $100$ days. Since the clouds move at $v \sim 10^4\;{\rm km}/{\rm s}$, the  size of the emission regions can be estimated to be $\sim 10^{-3}\pcs$~\cite{He:2018snd}. 

Although the size of individual emission regions can be constrained by the observed time scales, regions relevant for different periods of flaring might be much further apart. However, the overall extent of $\gamma$-ray and radio emissions can be determined to be $\mathcal{O}(\pc)$ using very-long-baseline interferometry (e.g.~\cite{Hodgson:2016ild}).
Even with the very conservative assumption that neutrinos can be emitted from a jet of galactic scale (kpc), we may assume that the neutrinos are essentially emitted from a point source when comparing to the typical distance to blazars.
The redshifts of observed blazars are expected to be in the  $z\sim (0.02-0.4)$ range~\cite{Padovani:2016wwn} which corresponds to proper distances between $\sim 100\Mpc$ and $2000\Mpc$.
The redshift of TXS 0506+056 is $0.3365 \pm 0.0010$~\cite{Paiano:2018qeq} corresponding to a proper distance of $1.741\pm0.005\Gpc$\footnote{Throughout the paper we use $\Omega_m = 0.315$ and $\Omega_\Lambda = 0.685$ for the density parameters for matter and the cosmological constant~\cite{Aghanim:2018eyx}, while we use $H_0=70$km/(s Mpc) for the Hubble constant.}.

The task of calculating the neutrino energy spectrum from a blazar from first principles is very challenging since we do not have information on the exact physical conditions such as the spectrum of the primary protons or the photon field. Instead, the neutrino spectrum is usually calculated by relating it to the low-energy synchrotron peak and the $\gamma$-ray spectrum. The neutrino energy flux can be described by~\cite{Padovani:2015mba}
\begin{equation}
  \label{eq:neutrino_spectrum}
  F_\nu (E_\nu) = F_0 \,E_\nu^{-s} \exp\left(-\frac{E_\nu}{E_0}\right).
\end{equation}
Here, $E_0$ is the peak energy of the neutrino spectrum which is in turn set by the peak in the observed synchrotron spectrum, the redshift, and the Doppler factor, which can all be derived from simulations of the SED. Typical values for $E_0$ are $10-20\PeV$. Based on self-consistent calculations of the neutrino spectra, the spectral index is $s = -0.35 \pm 0.12$~\cite{Petropoulou:2015upa}. Finally, the normalisation is given by
\begin{equation}
  \label{eq:nunorm}
  F_0 = \frac{F_{\nu, \rm tot} \, E_0^{s-1}}{\int_0^\infty dx \,x^{-s}\, e^{-x}}\,, 
\end{equation}
where the total neutrino flux is related to the $\gamma$-ray flux: $F_{\nu, \rm tot} = Y F_{\gamma}(E_\gamma>10^5 \GeV)$. Typical values of $Y$ range from 0.1 to 2~\cite{Petropoulou:2015upa}.
The spectrum in \cref{eq:neutrino_spectrum} has a sharp cutoff above $E_0 \sim 10\PeV$ due to exponential suppression. 
At lower energies (below $E_\nu=0.5\PeV$) blazars are only supposed to account for $\sim 10 \%$ of the IceCube neutrinos \cite{Padovani:2015mba, Padovani:2016wwn, Murase:2018iyl}. As it turns out, the neutrino associated with TXS 0506+056 is at the lower border of this window having a most probable energy of $290\TeV$ with a 90\% confidence interval from $183\TeV$ to $4.3\PeV$~\cite{IceCube:2018dnn}\footnote{This is assuming the spectral index $-2.13$. Assuming $-2.0$, the central value is $311\TeV$, and the limits are $200\TeV$ and $7.5\PeV$.}. For the $13\pm 5$ events found between September 2014 and March 2015 in the direction of TXS 0506+056, the individual neutrino energies are not reconstructed~\cite{IceCube:2018cha}. However, those that contribute most to the excess in the likelihood analysis have most probable energies of $10-100\TeV$ with confidence intervals covering at least an order of magnitude. Here, the large uncertainty in the neutrino energy stems from taking into account that observed muons were produced at an unknown distance from the detector (which is a defining property of the TGM \cite{Aartsen:2016xlq} dataset). It is worthwhile to point out that a large discrepancy between the energy deposited in the detector and the true neutrino energy also occurs in neutral current interaction of neutrinos of all flavors, where a final state neutrino can carry a significant portion of energy out of the detector \cite{Aartsen:2013vja}. The working hypothesis of our analysis (presented in \cref{sec:flavor_ratio}) is that there will be a number of high-energy neutrino events from blazars whose energy can be reconstructed with a precision of 10\%-15\% percent. This is achievable in case of charged current interaction of muon or electron neutrinos inside the detector volume. In such cases, the incoming neutrino energy practically matches the energy deposited in the detector which can indeed be determined with the precision mentioned above \cite{Aartsen:2013vja,Kopper:2017zzm}. 
High-energy neutrino events with such energy precision have already been measured: in the recently published 6-year HESE data \cite{Aartsen:2017mau} there are 22 muon tracks and 58 showers (latter containing both neutral and charged current events). 
The number of these events will increase in the forthcoming years, and will be accumulated at higher frequency when IceCube-Gen2 and KM3NeT-ARCA are constructed.
It is relevant to point out that the angular resolution of muon track events is $\lesssim 1^\circ$, whereas for showers the corresponding value is around $15^\circ$. However, with IceCube Upgrade, the resolution could improve to $0.1^\circ-0.2^\circ$ and $3^\circ-5^\circ$ respectively~\cite{ICupgrade2}. In the case of blazars, the time correlation between the duration of a given flaring episode and the arrival time of shower events gives additional identifying power. Such an analysis involving time information is currently being done by the IceCube collaboration for track events, and a similar study involving shower events is planned. In conclusion, 
it seems feasible to measure high energy neutrinos, arriving from blazars, of different flavors with reasonably small energy uncertainty. 

\section{Flavor Ratios}
\label{sec:flavor_ratio}
\noindent
In this section we will employ one of the leading methods for constraining new physics scenarios with IceCube data: the flavor composition of active neutrinos at Earth. As already discussed in \cref{sec:blazar_astro}, we focus on high-energy neutrinos from identified blazars such that the initial flavor composition and baseline are known. As demonstrated in the previous section, the baseline uncertainties arising from the finite size of individual emission regions as well as the separation between different emission regions in a blazar are much smaller than the average blazar distance from Earth and hence negligible. 

Neutrino flavor composition at Earth is independent of the baseline in a standard 3-flavor framework. This is because the oscillation length $L_\text{osc}= 4\pi E_{\nu}/\Delta m^2$, for mass squared differences between the known neutrino mass eigenstates ($\Delta m_\text{solar}^2\simeq 7.4\times 10^{-5}\, \text{eV}^2,\,\Delta m_\text{atm}^2\simeq2.5 \times 10^{-3}\, \text{eV}^2$) and neutrino energies $E_\nu\sim 0.1-10$ PeV, is much smaller than the baseline. Hence, the oscillations effectively average across such distances leaving no baseline dependence in the survival and transition probabilities. 
This is also the case in models featuring light sterile neutrinos \cite{Abazajian:2012ys} with $\Delta m^2 \sim \mathcal{O}(1) \,\text{eV}^2$ as well as non-standard neutrino interactions \cite{Miranda:2015dra}. However, there are cases in which baseline dependence is not washed out when neutrinos propagate over large distances. Two well-known representative scenarios are pseudo-Dirac neutrinos \cite{Kobayashi:2000md,Beacom:2003eu,Esmaili:2012ac} and 
neutrino decay \cite{Beacom:2002vi,Baerwald:2012kc,Bustamante:2016ciw}. While this section is devoted to the testability of pseudo-Dirac neutrinos from observing neutrino flavor compositions at Earth, we also refer to neutrino decay  in \cref{app}.

\subsection{Model and Methods}
\label{subsec:model_methods}
In order to generate non-vanishing active neutrino masses it is common to introduce right-handed neutrinos. The famous example is the type-I seesaw mechanism \cite{Goran,Minkowski,GellMann:1980vs,Yanagida:1979as} where small neutrino masses are realized from typically very large Majorana masses, $M_N$, of right-handed neutrinos. 
Here we consider a complementary option, dubbed the pseudo-Dirac scenario, 
where the neutrino mass term is dominantly of the Dirac type with a very small contribution from Majorana mass terms.
This yields three pairs of almost degenerate light states when three right-handed fields are introduced. The strongest constraint on the level of degeneracy between active and sterile components in each pair arises from solar neutrino measurements and yields $\Delta m^2 \lesssim 10^{-12}\,\text{eV}^2$ \cite{deGouvea:2009fp}. With the high-energy neutrinos traveling across astrophysical distances, even smaller values can be probed. In this section we present results for $\Delta m^2\lesssim 10^{-15}\,\text{eV}^2$.

The transition probability between active neutrino flavors $\alpha$ and $\beta$ reads \cite{Esmaili:2012ac}
\begin{align}
P_{\alpha\beta}=\frac{1}{4}\, \bigg|\sum_{i=1}^{3} U_{\alpha i} \left\{\text{Exp}\left[i \Phi_i^+\right]+\text{Exp}\left[i \Phi_i^-\right]\right\}
U_{\beta i}^*\bigg|^2,
\label{eq:Palphabeta}
\end{align}
where $U$ is the PMNS matrix. The phase $\Phi_i^+$ ($\Phi_i^-$) is associated to the active (sterile) component in the $i$-th pair and reads  \cite{Esmaili:2012ac}
\begin{align}
\Phi_i^\pm= \int_t^{t_0} dt'\, \bigg[(m_i^\pm)^2 + k^2 \left(\frac{a(t_0)}{a(t')}\right)^2\bigg]^{1/2} \simeq 
\int_t^{t_0} dt' \frac{k}{a(t')}  + \frac{(m_i^\pm)^2}{2} \int_t^{t_0} dt'\, \frac{a(t')}{k}\,,
\label{eq:Phi}
\end{align}
where $a$ is a scale factor and $k$ is the neutrino momentum, while $t_0$ and $t$ denote neutrino emission and detection time, respectively. In order to obtain the second equality, we defined $a(t_0)=1$ and used $(1+x^2)^{1/2}\simeq 1+x^2/2$ for $x\equiv m_i^\pm a(t')/k\ll 1$.

If the distances involved in the neutrino propagation were comparatively small, $a$ would not deviate from $1$ and \cref{eq:Palphabeta} would reduce to \cite{Beacom:2003eu,Esmaili:2009fk}
\begin{align}
P_{\alpha\beta}=\frac{1}{4}\, \bigg|\sum_{i=1}^{3} U_{\alpha i} \left\{\text{Exp}\left[\frac{i (m_i^+)^2 L}{2 E_\nu}\right]+\text{Exp}\left[\frac{i (m_i^-)^2 L}{2 E_\nu}\right]\right\}
U_{\beta i}^*\bigg|^2\,,
\label{eq:P_short}
\end{align}
where $L$ and $E_\nu$ are the baseline and neutrino energy, respectively, whereas the mass 
of the active (sterile) component in the i-th pair is denoted $m_i^+$ ($m_i^-$). This formula, however, is not applicable for astrophysical neutrinos from blazars whose redshift distribution peaks around $z\sim 1$. By employing $a(t)=1/(1+z)$ and the relation 
$|dz/dt|=H_0 (1+z) \sqrt{\Omega_m (1+z)^3 + \Omega_\Lambda}$
with which the time integral can be transformed into an integral over redshift $z$, 
the difference between phases given in \cref{eq:Phi} can be written as
\begin{align}
\Delta \Phi_i = \Phi_i^+ -\Phi_i^-= \frac{\Delta m_i^2}{2 E_\nu} \frac{c}{H_0} \int_0^z \frac{dz'}{(1+z')^2 \sqrt{\Omega_m (1+z')^3 + \Omega_\Lambda}}\,.
\end{align}
Here, we use the abbreviation $\Delta m_i^2 = (m_i^+)^2 - (m_i^-)^2$ for states in the i-th pair. After averaging over terms containing mass squared differences of the order of $\Delta m_\text{solar}^2$ and  $\Delta m_\text{atm}^2$, one finds that the flavor transition probability depends on $\Delta \Phi_i$~\cite{Esmaili:2012ac}:
\begin{align}
P_{\alpha\beta} = \sum_{i=1}^3 |U_{\alpha i}|^2 \,|U_{\beta i}|^2\, \cos^2\left(\frac{\Delta \Phi_i}{2}\right)\,.
\label{eq:P_used}
\end{align}
For the flavor ratio analysis, it is crucial to correctly infer the possible values of the
last factor in \cref{eq:P_used}. Clearly, in the limit  $\Delta \Phi_i\to 0$ this factor yields $1$, and there is no effective difference with respect to the standard case with three active neutrinos. In a similar fashion, if  $\Delta \Phi_i \gg 2 \pi$ for all three generations, the factor averages to $1/2$ leaving no observable effects in the flavor ratio analysis. In the following, we are primarily interested in values of $\Delta m_i^2$ for which $\cos^2\left(\Delta \Phi_i /2\right)$ has a non-trivial impact on the flavor composition.

Blazar sources are numerous and $\cos^2\left(\Delta \Phi_i /2\right)$ should, in case the contributing sources are not identified,  be weighted by the redshift distribution $R(z)$ (adopted from Ref. \cite{Padovani:2015mba}, Fig. 3). In order to compare averaged and non-averaged results, we follow Ref. \cite{Esmaili:2012ac} and define effective
value of $\cos^2\left(\Delta \Phi_i /2\right)$, henceforth dubbed the suppression factor\footnote{The motivation for this name stems from the comparison with the standard relation where only three active neutrinos are involved. There, $\cos^2\left(\Delta \Phi_i /2\right)$ is not present and 
hence for any $\cos^2\left(\Delta \Phi_i /2\right)\neq 1$ we observe a ``suppression".} as 
\begin{align}
S_\text{eff}\,(\Delta m_i^2, E_\nu)&=\frac{\int_{z=0}^\infty dz' \, R(z') \,dV_c/dz' \,\left[\int dE_\nu' \cos^2\left(\frac{\Delta \Phi_i (E_{\nu}',z')}{2}\right) G(E_\nu,E_\nu')\right] \mathcal{F}(E_\nu,z')/d_c(z')^2}
{\int_{z=0}^\infty dz' \, R(z') \,dV_c/dz'\, \mathcal{F}(E_\nu,z')/d_c(z')^2}  \nonumber \\ 
& \equiv \frac{\text{Num}(\Delta m_i^2,E_\nu)}{\text{Den}(E_\nu)}\,.
\label{eq:Seff}
\end{align}
Here, $dV_c$ and $d_c$ are the comoving volume element and comoving distance, respectively, $\mathcal{F}$ is the neutrino flux from blazars which we infer from the energy flux given in \cref{eq:neutrino_spectrum}\footnote{The parameter $s$ in \cref{eq:neutrino_spectrum} is fixed to the best fit value of $-0.35$ \cite{Padovani:2015mba}. The energy $E_0$ depends on redshift over which we integrate in \cref{eq:Seff}. We fix  $E_0=17.5\,\text{PeV}/(1+z)^2$ \cite{Padovani:2015mba}, where  the value of the Doppler factor and synchrotron frequency are taken to be $10$ and $10^{16}$ Hz, respectively.}. We account for the energy smearing in the detector by convoluting the oscillating term in the numerator with a gaussian, $G$, that has a mean value centered at the true neutrino energy $E_\nu$ and standard deviation which equals $0.1 E_\nu$. The latter is a realistic energy uncertainty, as discussed in \cref{sec:blazar_astro}. We have also checked that the neutrino coherence is preserved for values of $E_\nu$, $z$ and $\Delta m_i^2$ under consideration. In what follows, the term in the square brackets will be denoted as $\langle \cos^2\left(\Delta \Phi_i (E_{\nu},z)/2\right) \rangle$ for brevity.

Given that IceCube can detect neutrino energies in a certain range $(E_\nu^\text{min}, E_\nu^\text{max})$, we also define the energy averaged suppression factor \cite{Esmaili:2012ac} as 
\begin{equation}
  \label{eq:1}
\overline{S}_\text{eff}(\Delta m_i^2) = \frac{\int_{E_\nu^\text{min}}^{E_\nu^\text{max}} dE_\nu\, \text{Num}(\Delta m_i^2,E_\nu)}{\int_{E_\nu^\text{min}}^{E_\nu^\text{max}} dE_\nu\, \text{Den}(E_\nu)} .
\end{equation}
Employing such averaging is appropriate for a comparison with current IceCube limits on flavor ratios as the latter are calculated from a set of neutrino events with different energies. 

Depending on whether an integral over energy and/or redshift is performed, $\cos^2\left(\Delta \Phi_i /2\right)$ in \cref{eq:P_used} is replaced by either $S_\text{eff}$ or $\overline{S}_\text{eff}$ in order to account for the finite range of energies and blazars at different redshifts that contribute to the signal.
Let us note that $S_\text{eff}$ and $\overline{S}_\text{eff}$ can also be defined for a single known blazar (or a set of blazars) by dropping the integral over redshift leaving only integration over energy smearing in the detector or integration over a full energy range. This will be employed in \cref{subsec:results} where we will focus on comparing scenarios with known and unknown baselines. 

After defining the suppression factors that enter in the transition probabilities, we can
relate flavor abundances at production and detection 
\begin{align}
\begin{pmatrix}
 X_{e} \\[0.2em]
    X_{\mu}   \\[0.2em]
      X_{\tau}  
    \end{pmatrix}  =
    \begin{pmatrix}
 P_{\nu_e \rightarrow \nu_e} & P_{\nu_e \rightarrow \nu_\mu} & P_{\nu_e \rightarrow \nu_\tau}\\[0.2em]
     P_{\nu_e \rightarrow \nu_\mu}   & P_{\nu_\mu \rightarrow \nu_\mu} & P_{\nu_\mu \rightarrow \nu_\tau} \\[0.2em]
      P_{\nu_e \rightarrow \nu_\tau} & P_{\nu_\mu \rightarrow \nu_\tau} & P_{\nu_\tau \rightarrow \nu_\tau} 
    \end{pmatrix} 
   \begin{pmatrix}
 X^{\text{in}}_{e} \\[0.2em]
    X^{\text{in}}_{\mu}   \\[0.2em]
      X^{\text{in}}_{\tau}  
    \end{pmatrix}\,,
    \label{eq:final_initial}
\end{align}
where $X_{\alpha}$ ($X^{\text{in}}_{\alpha}$) represents the final (initial) relative abundance of high-energy neutrinos of flavor $\alpha$ in the full set of observed neutrinos. As already discussed in \cref{sec:blazar_astro}, the flavor ratios at production are expected to be
 $(X^{\text{in}}_e:X^{\text{in}}_\mu:X^{\text{in}}_\tau) = (1:2:0)$ for blazars.
 
Let us briefly mention how the elements of PMNS matrix, which enter \cref{eq:P_used} and 
 consequently \cref{eq:final_initial}, are generated. Following Ref. \cite{Rasmussen:2017ert} we define $\chi^2$ as
  \begin{align}
 \chi^2 = \sum_{ij=12,13,23} \left(\frac{\sin^2 {\theta_{ij}}-(\sin^2 {\theta_{ij}})^\text{bfp}}{\sigma_{\sin^2 {\theta_{ij}}}}\right)^2,
 \label{eq:chi2}
 \end{align}
where the best fit point (denoted bfp) and standard deviation for all three mixing angles are adopted from Ref.~\cite{Esteban:2018azc} assuming normal mass ordering (the results are qualitatively similar for inverted mass ordering). We require $\chi^2$
to be within the 99\% CL, i.e. $\chi^2<11.83$. The Dirac CP phase $\delta$ is taken to be random in our simulations since robust statistical results for this parameter are lacking. The parametrization of the employed PMNS matrix in terms of mixing angles and Dirac CP phase
(Majorana phases are not observable) may be found in Ref. \cite{PDG}.

\subsection{Results}
\label{subsec:results}

\begin{figure}
	\centering
	\subfloat[\label{fig:1left}]{\includegraphics[scale=0.29]{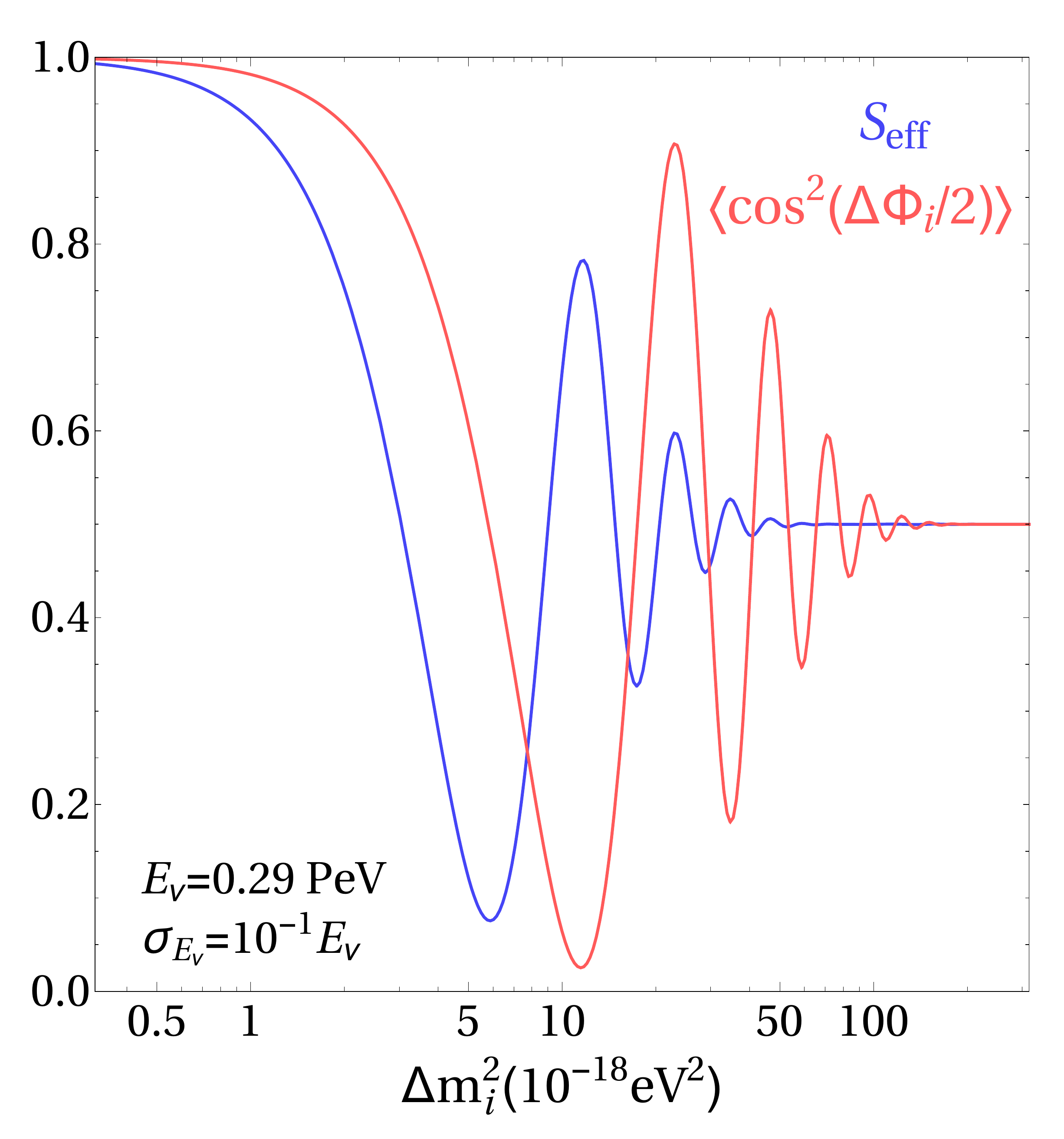}}
	\hspace{4em}%
	\subfloat[\label{fig:1right}]{\includegraphics[scale=0.29]{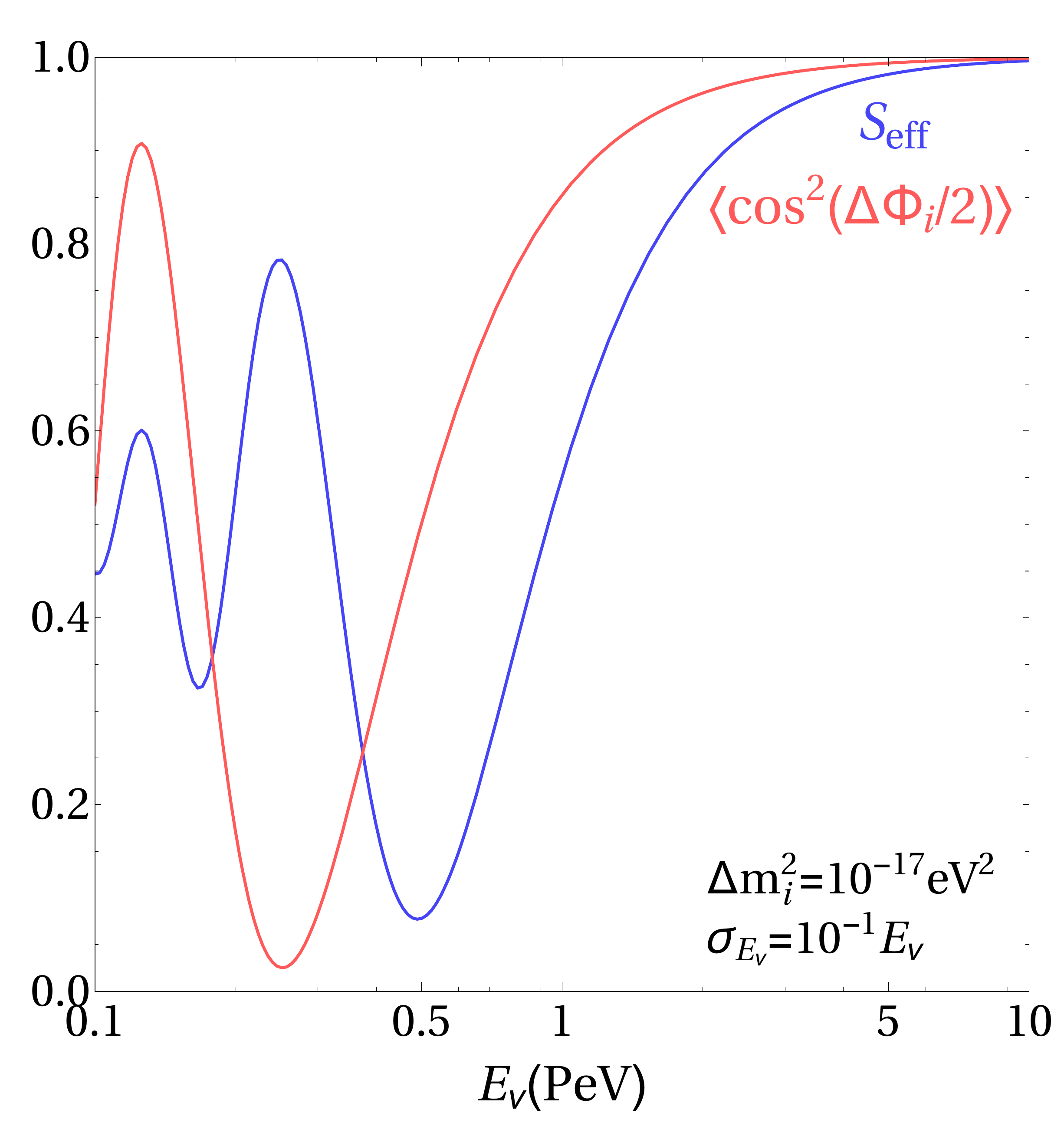}}
	\caption{(a) The blue curve shows $S_\text{eff}$ calculated using \cref{eq:Seff}, whereas the red curve represents $\langle \cos^2\left(\Delta \Phi_i/2\right) \rangle$ 
for a single blazar at a distance of $z=0.3365$.	
	Both curves are shown as a function of $\Delta m_i^2$. (b) The same quantities as a function of neutrino energy for fixed $\Delta m_i^2=10^{-17}\,\text{eV}^2$.}
	\label{fig:SeffPD}
\end{figure}

In this section we present results of the flavor composition analysis obtained using tools introduced in \cref{subsec:model_methods}. We start off by comparing $S_\text{eff}$ (blue curve in \cref{fig:1left} calculated by weighting over blazar redshift distribution; see again \cref{eq:Seff}) and $\langle \cos^2\left(\Delta \Phi_i (E_{\nu},z_B)/2\right) \rangle$, both evaluated for a fixed energy of $E_\nu=0.29$ PeV which matches the most probable energy of the recently observed neutrino from TXS 0506+056. We show $\langle \cos^2\left(\Delta \Phi_i (E_{\nu},z_B)/2\right) \rangle$ as a red curve for which we adopt the  measured redshift of TXS 0506+056, $z_B=0.3365$. In both curves we observe the expected behavior for very small and large values of $\Delta m_i^2$. A small value of $\Delta m_i^2$ is equivalent to the standard 3-flavor scenario where the active-sterile oscillations do not develop. A large $\Delta m_i^2$ gives averaging of the trigonometric function, i.e. picking up a factor of $1/2$. Note that the figure is valid for each of the three pairs of active and sterile neutrinos, scanned in the relevant range of $\Delta m_i^2$.

The two curves are clearly out of phase and this is because the redshift distribution of blazars, $R(z)$, peaks around $z\sim 1$, differing from $z_B\sim 0.34$. It is much more likely to observe $\gamma$-rays from not so distant blazars at $z<1$ (see in particular Fig. 3 in Ref. \cite{Padovani:2015mba}) and therefore the presented comparison is qualitatively robust when confronting a single identified source with neutrinos from a distribution of unknown blazars. By considering all relevant values of $\Delta m_i^2$, we see that the maximal suppression is achieved for a single source (red curve) at a value of $\sim 10^{-17} \, \text{eV}^2$. 
The shown $S_\text{eff}$ does not reach as small values since it is smeared over a distribution of redshifts whereas the suppression factor for the single source is only smeared over the energy resolution of the detector.
Here we see the first advantage stemming from knowing a location of the source: The accessible new physics parameter space opens up since the deviation of the suppression factor from the value of $1$ should be regarded as the deviation from the standard case involving only three active neutrino states. This means that the total parameter space of possible flavor compositions in scenarios where the sources are known is larger. 
In particular, in several identified regions such as for $\Delta m_i^2\simeq (8-16)\times 10^{-18}\,\text{eV}^2$, we observe that the red line is below the blue one, indicating a parameter range where the suppression factor departs more strongly from the standard case for $z_B$ than for the redshift distribution. At $\sim 11\times 10^{-18}\,\text{eV}^2$ the difference is maximal and yields $\sim 0.8$. There is a caveat though: Although, seemingly, the pseudo-Dirac model can be robustly probed using aforementioned arguments by performing a flavor analysis, note that the value of $E_\nu$ used for producing the figure is fixed to $E_\nu=0.29$ PeV, and hence a number of events around that (or any other) energy and similar redshifts is required. Alternatively, one can combine different energies and redshifts, but similar $\Delta \Phi_i$ in the same analysis. This will be achievable with large statistics in case of which the IceCube collaboration may perform the analysis of the flavor composition as a function of neutrino energy. Needless to say, the positions of the regions with maximal suppression factor shift if different neutrino energies are considered; in other words the values of $\Delta m_i^2$ with the strongest testability (provided enough events in a given bin) are altered for different values of $E_\nu$.
 
It is worth noticing that there is an uncertainty in the determination of $S_\text{eff}$. By comparing our integrated results for $\gamma$-ray bursts to Ref. \cite{Esmaili:2012ac}, we find a deviation of $\Delta S \sim 0.05$. This is due to the sampling of $\mathcal{O}(100)$ \emph{unidentified} sources, employed in Ref. \cite{Esmaili:2012ac}. This influences the robustness of present constraints in flavor ratio analysis for the considered model. There is no such problem for \emph{identified} sources as the redshift can be reliably taken into account.

\begin{figure}
	\centering
	\includegraphics[scale=0.29]{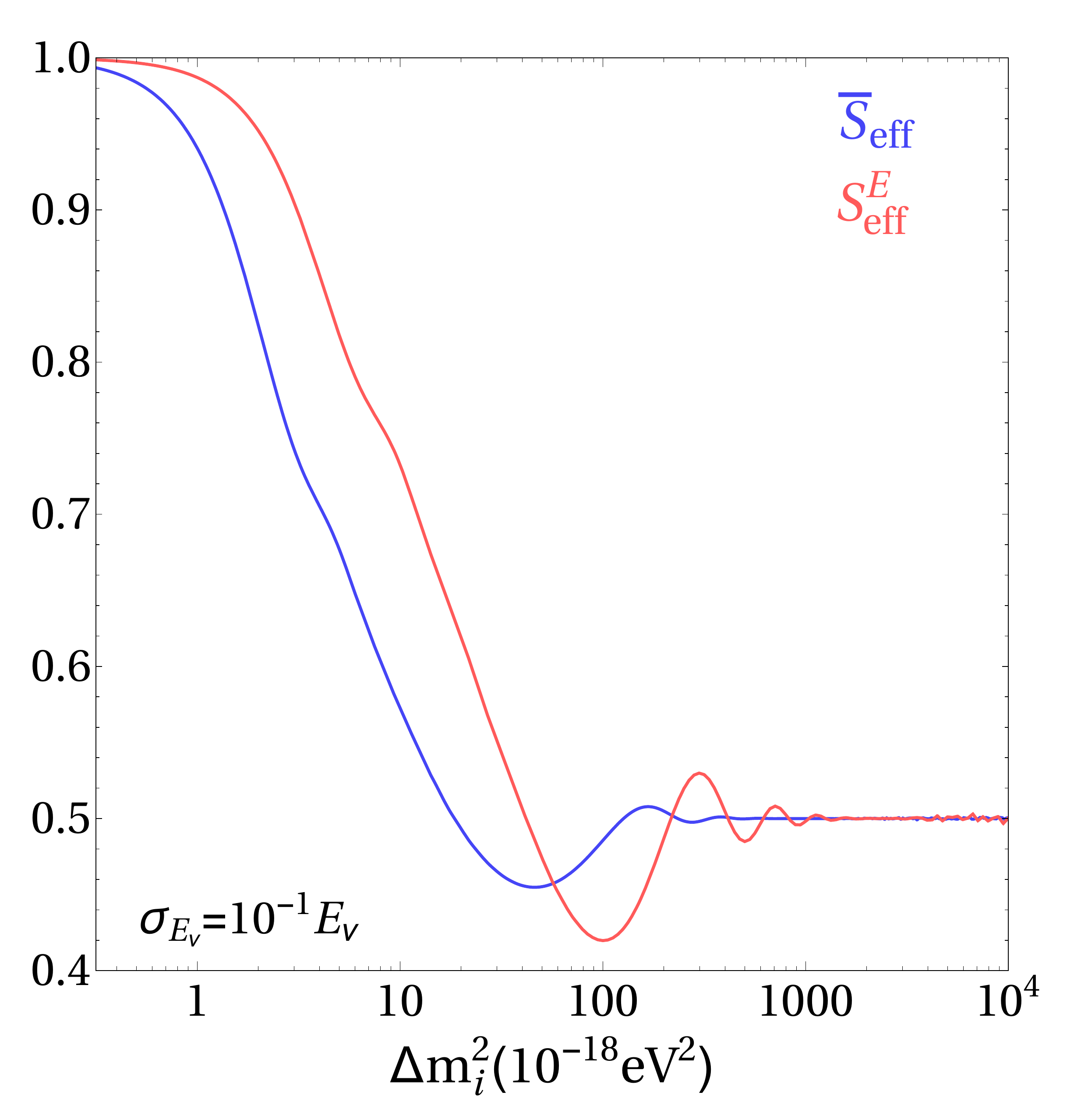}
	\caption{$\overline{S}_\text{eff}$ and energy-averaged value of $\langle \cos^2\left(\Delta \Phi_i (E_\nu,z_B=0.3365)/2\right) \rangle$  are shown in blue and red, respectively. The integrals over neutrino energy are performed in range $(E_\nu^\text{min},E_\nu^\text{max})=(0.1\,\text{PeV}, 5\,\text{PeV})$.}
	\label{fig:BARSeffPD}
\end{figure}

In order to provide a more complete picture, \cref{fig:1right} shows  $S_\text{eff}$ and $\langle \cos^2\left(\Delta \Phi_i (E_\nu,z_B)/2\right) \rangle$ for $\Delta m_i^2=10^{-17}\,\text{eV}^2$ as a function of neutrino energy. At the energy of $0.3$ PeV, the discrepancy between the curves is maximal (as can also be inferred from \cref{fig:1left}), but becomes milder at higher values of $E_\nu$.

To capture the energy-averaged effect, we show $\overline{S}_\text{eff}$ (blue) in \cref{fig:BARSeffPD} as well as the energy-averaged value of $\langle \cos^2\left(\Delta \Phi_i (E_\nu,z_B)/2\right) \rangle$ for  $z_B=0.3365$ (red; denoted as $S_\text{eff}^E$). We took $(E_\nu^\text{min},E_\nu^\text{max})=(0.1\,\text{PeV}, 5\,\text{PeV})$, but have also checked that varying $E_\nu^\text{max}$ by a factor of few does not yield qualitatively different results. As can be inferred from the figures, the energy averaged suppression is not as large as the suppression shown in \cref{fig:SeffPD}. This leads us to the conclusion that the energy dependent analysis of the flavor composition is more promising for probing this model. To achieve that, as already argued, one requires a large number of events in at least one of the energy bins. This could nevertheless be possible, especially in light of proposed IceCube detector upgrades.
 From \cref{fig:BARSeffPD} we infer that the maximal suppression for the distribution of unknown sources (a single known source) is $0.45$ ($0.42$). This still shows a minor but existing advantage in favor of the scenario where the redshift of the blazar(s) is known. Note that the minima of both curves in \cref{fig:BARSeffPD} lie at larger values of $\Delta m_i^2$ when compared to the first oscillation minima in \cref{fig:1left}. This is because $E_\nu$ in \cref{fig:1left} is fixed to a value smaller than $1$ PeV. For $E_\nu\sim \mathcal{O}(1)$ PeV, the extrema would be shifted to larger $\Delta m_i^2$, and such energies indeed dominate in the $(E_\nu^\text{min},E_\nu^\text{max})$ range employed for producing \cref{fig:BARSeffPD}.

\begin{figure}
	\centering
	\subfloat[\label{fig:triangle-x}]{\includegraphics[width=7.5cm,height=7.5cm]{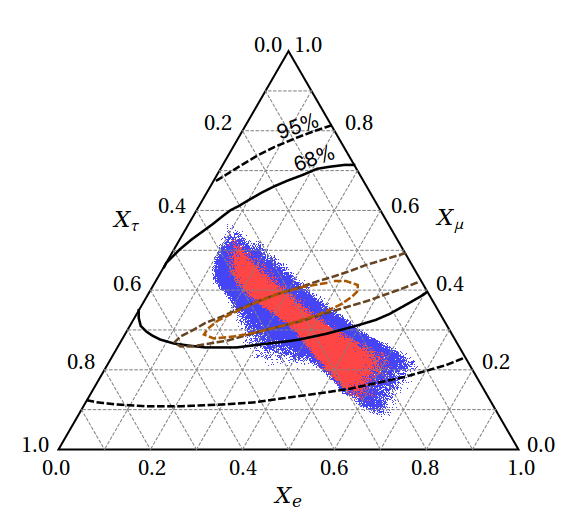}}
	\hspace{4em}%
	\subfloat[\label{fig:x}]{\includegraphics[scale=0.29]{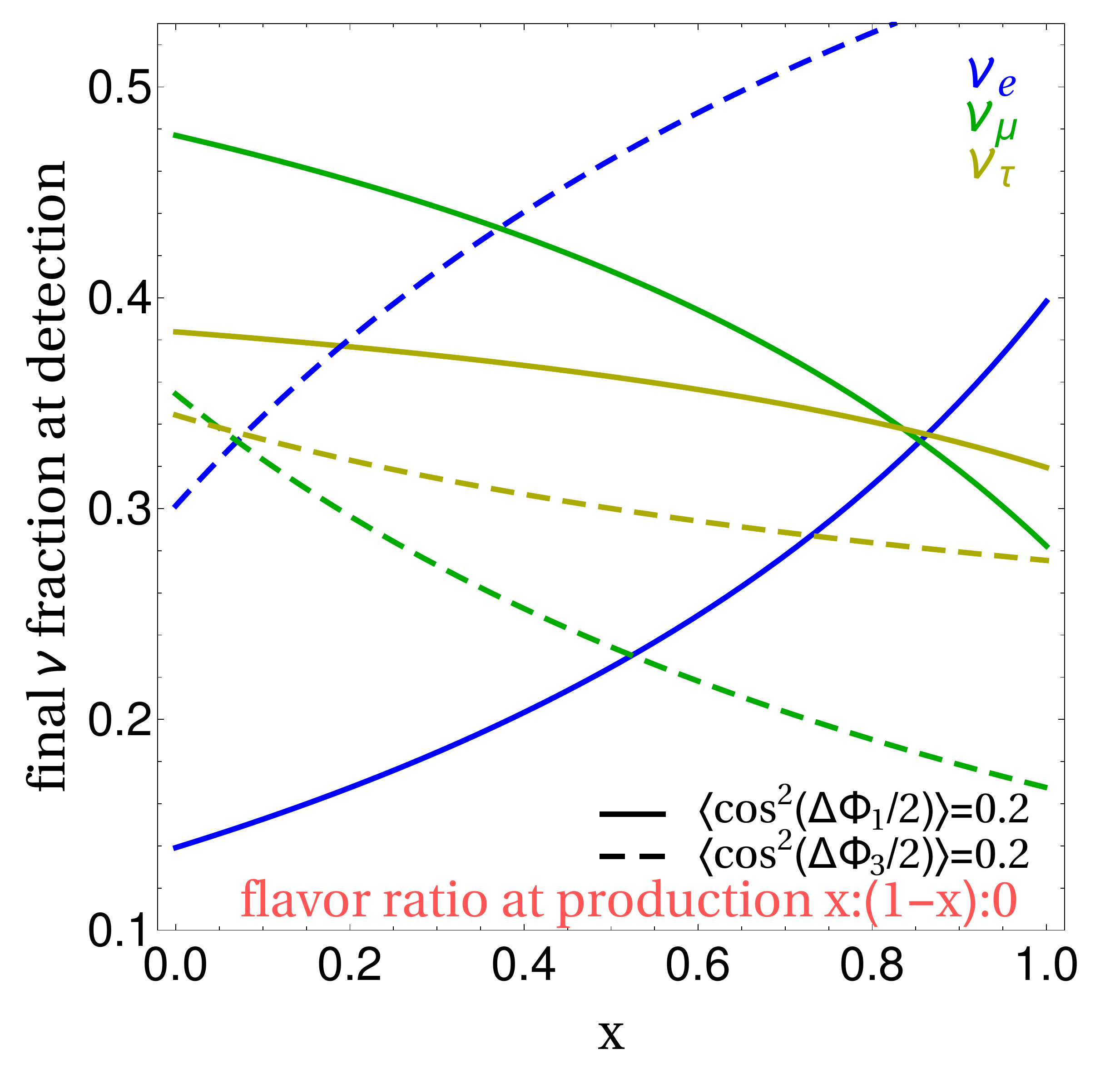}}
	\caption{(a) Each of the sides of a ternary plot represents the faction of 
	neutrinos of a given flavor. The flavor composition at the source is taken as $\left(x:(1-x):0\right)$, where $x$ is randomly drawn $10^5$ times in the range between $0$ and $1$.
 To obtain the red curve we consider suppression factors between $0.42$ and $1$ which is the accessible range for a single source when the energy average is performed (see \cref{fig:BARSeffPD}). The blue range corresponds to the union of three sets, each of which 
 obtained by varying suppression factor for $i$-th active-sterile pair in $(0.025,0.3)$ range and fixing the two remaining ones to $1$. For each value of neutrino energy there are $\Delta m_i^2$ for which the suppression can reach such values. The shown regions are compared with present HESE limits (black) and future (dark orange, dark brown) sensitivity of upgraded IceCube detector. (b) Fraction of each flavor at Earth as a function of $x$. The figure confirms what is already visible in (a), namely that  the electron and muon neutrino fractions  change more strongly than the tau fraction  in the viable parameter range.
 }
	\label{fig:FTx}
\end{figure}

\begin{figure}
	\centering
\subfloat[\label{fig:triangle-120}]{\includegraphics[width=7.5cm,height=7.5cm]{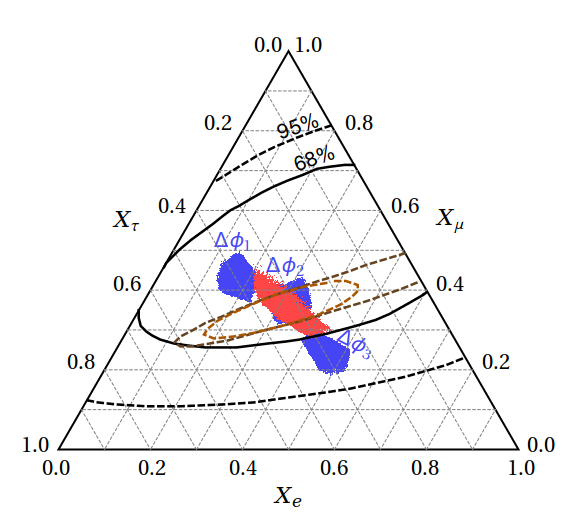}}
	\hspace{4em}%
	\subfloat[\label{fig:120}]{\includegraphics[scale=0.29]{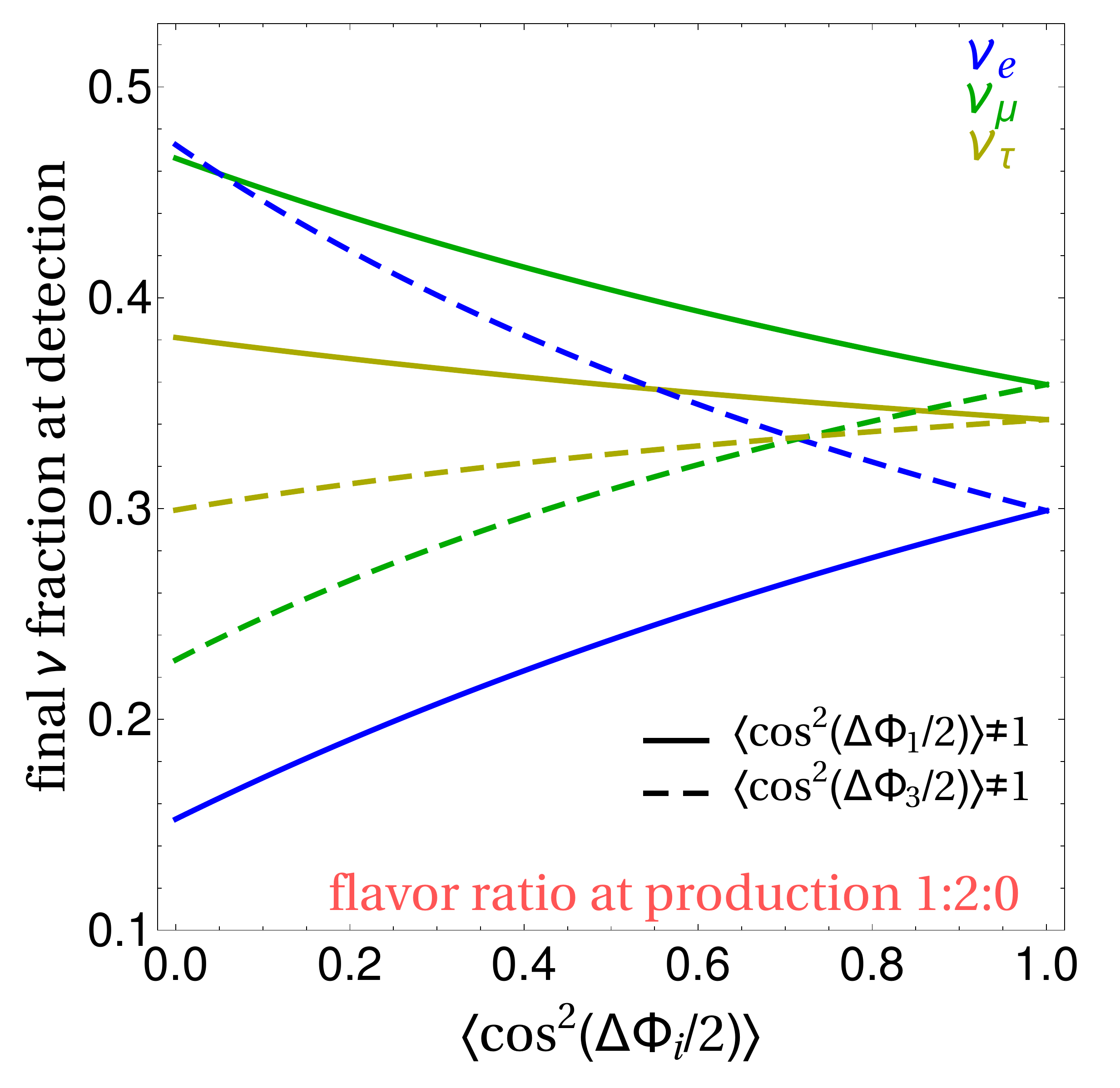}}
	\caption{(a) Same as \cref{fig:triangle-x}, only for fixed initial flavor ratio 
	$(\nu_e:\nu_\mu:\nu_\tau) = (1:2:0)$ which is realistic for blazars. The phase that is varied is indicated for each blue region. (b) Flavor compositions at Earth as a function of $\langle \cos^2\left(\Delta \Phi_i/2\right) \rangle$.}
	\label{fig:FT120}
\end{figure}

In \cref{fig:triangle-x,fig:triangle-120} we present our results in terms of ternary plots where each of three sides of a triangle represents a relative portion of a given flavor, $X_{\alpha}$ ($\alpha$=e,\,$\mu$,\,$\tau$), in a set of high-energy neutrinos detected at Earth. 
In \cref{fig:triangle-x} we consider the most general astrophysical scenario with 
$(X^{\text{in}}_e:X^{\text{in}}_\mu:X^{\text{in}}_\tau) = (x:1-x:0)$, where $x$ is generated randomly $10^5$ times with a value between $0$ and $1$. In this way we are simultaneously taking into account several astrophysical high-energy neutrino production mechanisms. Namely, for $x=0$ we have $(0:1:0)$ which corresponds to the case where muons from pion decays rapidly lose energy and do not yield two additional high-energy neutrinos in the decay. For $x=1/3$ we have the pion decay case $(1/3:2/3:0)$, whereas the neutron decay source corresponds to $x=1$ $(1:0:0)$. To obtain the flavor compositions at Earth, we use \cref{eq:final_initial} with the transition probability $P_{\alpha\beta} = \sum_{i=1}^3 |U_{\alpha i}|^2 \,|U_{\beta i}|^2\, \mathcal{S}_i$, where $\mathcal{S}_i$ corresponds to the suppression factor for an individual source with known redshift and for the $i$-th active-sterile pair. These suppression factors are therefore analogues to ${S}_\text{eff}$ and $\bar{S}_\text{eff}$ which we defined for unknown distributions (see \cref{subsec:model_methods}).
The values of $\mathcal{S}_i$ can be inferred from \cref{fig:SeffPD,fig:BARSeffPD} (red curves).
As reported above, the maximal energy-averaged suppression gives $\mathcal{S}_i=0.42$. The red region in \cref{fig:triangle-x} corresponds to the maximal accessible range 
of flavor compositions at Earth in case energy averaging is employed. To obtain this region we randomly generated $\mathcal{S}_i$ $10^5$ times (for all three pairs simultaneously) between $0.42$ and $1$. This region can be compared to current IceCube flavor composition constraint obtained by using the 6-year HESE sample only \cite{HESEflavor} (black solid and dashed lines represent 68\% and 95\% CL, respectively). There is a significant portion of parameter space that exceeds the 68\% CL line. The testability will further improve with larger exposure and the upgrade of the IceCube detector. The dark brown (dark orange) curve represents the sensitivity of IceCube after 12 years of running (12 years of running+IceCube Upgrade\footnote{Note that these future projections include both HESE and TGM sample; HESE-only sensitivity with IceCube Upgrade is not currently available.} \cite{ICupgrade}). However, mapping from the excluded regions to an exclusion in $\Delta m_i^2$ is complicated by degeneracies between the unknown electron flavor fraction at the source, $x$, and $\Delta m_i^2$.
To generate the blue region shown in \cref{fig:triangle-x}, we revisit \cref{fig:1left}
from where we infer that a single source (red line) exhibits stronger suppression than the  distribution of sources for a substantial range of 
$\Delta m_i^2$. We focus on the region around $\Delta m_i^2\sim 10^{-17}\,\text{eV}^2$ (strongest single source suppression) where the suppression factor lies in the range $\mathcal{S}_i=(0.025,0.3)$. Such suppression values are achievable and experimentally testable if the IceCube detector discovers several neutrinos of similar energy from a known blazar or several neutrinos from different known blazars with the same $\Delta \Phi_i$. This phase does not need to correspond to the benchmark point with $E_\nu=0.29$ PeV. If it corresponds to a larger energy, e.g. $E_\nu\sim$ few PeV, the corresponding value of $\Delta m_i^2$ for which such suppression factors are expected shifts towards $10^{-16}\,\text{eV}^2$. We randomly generated $\mathcal{S}_i$ between $0.025$ and $0.3$ for the $i$-th pair while fixing $\mathcal{S}$ to $1$ for other two pairs. The procedure is repeated for each of the three pairs and the three regions are jointly shown as a blue region in \cref{fig:triangle-x}. This corresponds to the case where neutrino oscillations between active and sterile neutrino occur in one pair, while for other pairs they do not develop.
We point out  that the accessible region would be even larger if we fixed $\mathcal{S}_i=1$ for one pair and varied two other values simultaneously in the $(0.025,0.3)$ range. However, this is a very specific case as it would require two out of three values of $\Delta m_i^2$ to be very similar\footnote{For neutrino decay, a similar coincidence is not required since a decay only happens once.}.
 For completeness, let us state that having all three $\Delta m_i^2$ of similar size would only yield a global suppression factor of the neutrino flux and such a case is not observable in flavor composition. 

 We observe from  \cref{fig:triangle-x} that the accessible range in the ternary plot features the largest variation in the electron and muon neutrino fraction. This is also visible in \cref{fig:x} where we show fractions of each flavor at Earth as a function of electron neutrino flavor fraction at the production, denoted $x$. The mixing angles and  Dirac CP phase are fixed to current best fit values for normal ordering \cite{Esteban:2018azc}. The solid (dashed) 
 curve corresponds to $\langle \cos^2\left(\Delta \Phi_1/2\right) \rangle=0.2$
($\langle \cos^2\left(\Delta \Phi_3/2\right) \rangle=0.2$), while the values for the other two pairs are taken to be $1$, i.e. oscillations are only present in one pair of active-sterile neutrinos. Indeed, we observe that growth in the electron neutrino fraction is accompanied with a similar drop in the muon neutrino fraction, while the fraction of tau neutrinos is almost $x$-independent.

The results presented in \cref{fig:triangle-x} cover many different initial flavor compositions, but for a blazar we expect the flavor ratio of emitted neutrinos to be $(X^{\text{in}}_e:X^{\text{in}}_\mu:X^{\text{in}}_\tau) = (1:2:0)$~\cite{Hummer:2010vx,Kelner:2006tc,Hummer:2010ai,Kashti:2005qa}. 
In \cref{fig:triangle-120} we redo the aforementioned analysis for the flavor composition appropriate for blazar neutrinos. When comparing blue and red regions between \cref{fig:triangle-x,fig:triangle-120}, one notices that those in the latter are smaller due to the more constrained flavor composition at the source. In fact, we see in \cref{fig:triangle-120} that the three blue regions, originating from fixing two of the suppression factors to $1$ and varying the remaining one, are fully separated (Each region is marked by the varied phase).
To exclude or confirm any of the three blue regions, a sample of blazar neutrinos that have energy and redshift corresponding to a narrow range of $\Delta \Phi_i$ and with statistics comparable to a 12-year HESE sample will be needed.
Again, we generally find that the flavor compositions in the accessible range differ in electron and muon flavor, whereas there is no large variation for tau neutrinos. 
This may be also seen in \cref{fig:120} where we show flavor compositions at Earth as a function of $\langle \cos^2\left(\Delta \Phi_i/2\right) \rangle$. 
The solid (dashed) curves correspond to varying $\Delta \Phi_1$ ($\Delta \Phi_3$), while the phase differences of the remaining two pairs are assumed to be much smaller, such that the oscillations between those states do not develop. When $\Delta \Phi_1$ ($\Delta \Phi_3$) is varied, electron fraction tends to increase (decrease) with $\langle \cos^2\left(\Delta \Phi_i/2\right) \rangle$, and for muon fraction it is vice versa. Expectedly, both curves corresponding to a given flavor intersect at $\langle \cos^2\left(\Delta \Phi_i/2\right) \rangle=1$, case in which the standard three flavor oscillation scenario is reproduced and the model of pseudo-Dirac neutrinos is not testable. The flavor composition at Earth  is then close to $(1:1:1)$. 
The simple dependence of the flavor compositions as a function of $\Delta \Phi_i$ suggests a different way to test the pseudo-Dirac model even if the statistics for one value of $\Delta \Phi_i$ is insufficient to make an exclusion in \cref{fig:triangle-120}.
By dividing events into groups of different $\Delta \Phi_i$, the shape of $X_\alpha(\langle \cos^2\left(\Delta \Phi_i/2\right) \rangle)$ can be tested, and neutrinos of all energies and redshifts can be used to constrain the model.

In summary, in this section we explored the flavor composition analysis of the pseudo-Dirac neutrino model (introduced in \cref{subsec:model_methods}). In ternary plots we presented the accessible range for a single blazar. We showed the case in which suppression factors are integrated over energy yielding an ``energy-averaged'' information as well as the case for neutrinos of a similar energy where, for particular values of mass squared differences, very large suppression factors are reached, making the flavor composition analysis particularly sensitive.

\section{Summary}
\label{sec:summary}
\noindent

The first association of a high-energy neutrino with a blazar suggests that  many more neutrinos in the future could be linked to known sources. While knowing the sources that yield the
most energetic neutrinos ever observed is a valuable information itself, 
there are plenty of more exciting properties to be learned about such distant objects.
In this work we take a complementary perspective and ask the following question: What can
be learned about neutrinos if (blazar) sources are known? In particular, we focus on beyond the Standard Model scenarios. We find that the flavor composition analysis from IceCube may serve as a powerful probe of models in which the baseline information is not washed out from the oscillation probabilities after neutrinos propagate over astrophysical distances. While this is not the case for a number of beyond the Standard Model extensions such as light sterile neutrinos and non-standard interactions, we infer that considerations of pseudo-Dirac neutrinos in light of emission from known sources is relevant. For this model, we have studied neutrino flavor transition probabilities by confronting scenarios with \emph{known} and \emph{unknown} sources. The results from the latter are less reliable since drawing from a theoretically postulated redshift distribution cannot map the real physical situation. We have presented our results in the form of ternary diagrams where fractions of each neutrino flavor at Earth are shown. 
As a benchmark, we took TXS 0506+056 at $z=0.3365$ and assumed a significant number of events with similar source redshift and energy, but better energy resolution. This is required before our calculated regions can be successfully confronted with observations. 
We presented both energy-averaged scenarios as well as the cases for a fixed energy. Both cases could be probed by IceCube in the future. The latter yields larger number of  possible flavor compositions when focusing on a certain range of model's parameters, i.e. the deviations from the standard 3-flavor oscillation result increase.
In conclusion, the baseline information, i.e. identification of high-energy neutrino sources is relevant when considering models where the oscillation length is comparable to the propagated distance.

\section*{Acknowledgements}
\noindent
We thank Carlos Arg\"uelles and Arman Esmaili for several very useful discussions.

\appendix
\section{Neutrino decay}
\label{app}

In \cref{subsec:results} we focused on the pseudo-Dirac neutrino model, but there are other scenarios where the baseline dependence may not be washed out in the course of propagation from the known (blazar) source to Earth. One such example is neutrino decay where the propagating distance is dependent on the neutrino lifetime. 
Following Ref. \cite{Baerwald:2012kc}, we infer that the neutrino transition probability
(which is for pseudo-Dirac neutrinos is given in \cref{eq:P_used}) for neutrino decay\footnote{We assume for simplicity invisible neutrino decays, i.e. no transitions between neutrino mass eigenstates.} reads

\begin{align}
P_{\alpha\beta} = \sum_{i=1}^3 |U_{\alpha i}|^2 \,|U_{\beta i}|^2\, \text{Exp}\left[-\frac{\kappa}{E_\nu} \frac{c}{H_0} \int_0^z \frac{dz'}{(1+z')^2 \sqrt{\Omega_\text{matter} (1+z')^3 + \Omega_\Lambda}}\right],
\label{eq:P_used_exp}
\end{align}
where $\kappa_i$ is the ratio of neutrino mass $m_i$ to the rest-frame lifetime $\tau_i$ of the mass eigenstate $\nu_i\, (i=1,2,3)$. The current leading bound on $\kappa_1$ stems from Supernova 1987a and yields $\kappa_1\lesssim 10^{-5}$ eV/s \cite{Baerwald:2012kc}. With astrophysical neutrinos such values cannot be probed, however leading constraints on the decay of the heavier two neutrino mass eigenstates could be set.

\begin{figure}
	\centering
	\subfloat[\label{fig:appleft}]{\includegraphics[scale=0.29]{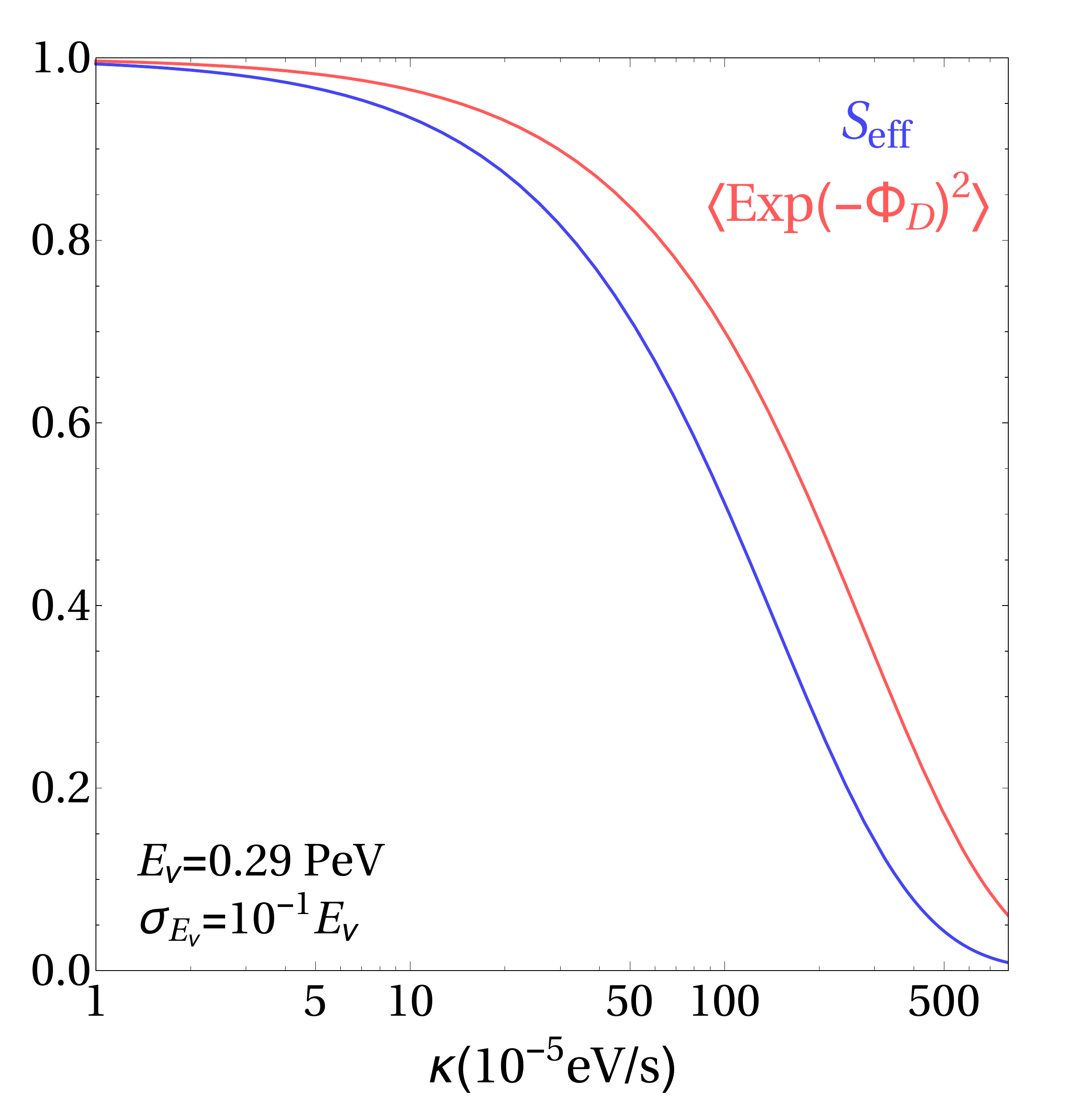}}
	\hspace{4em}%
	\subfloat[\label{fig:appright}]{\includegraphics[scale=0.29]{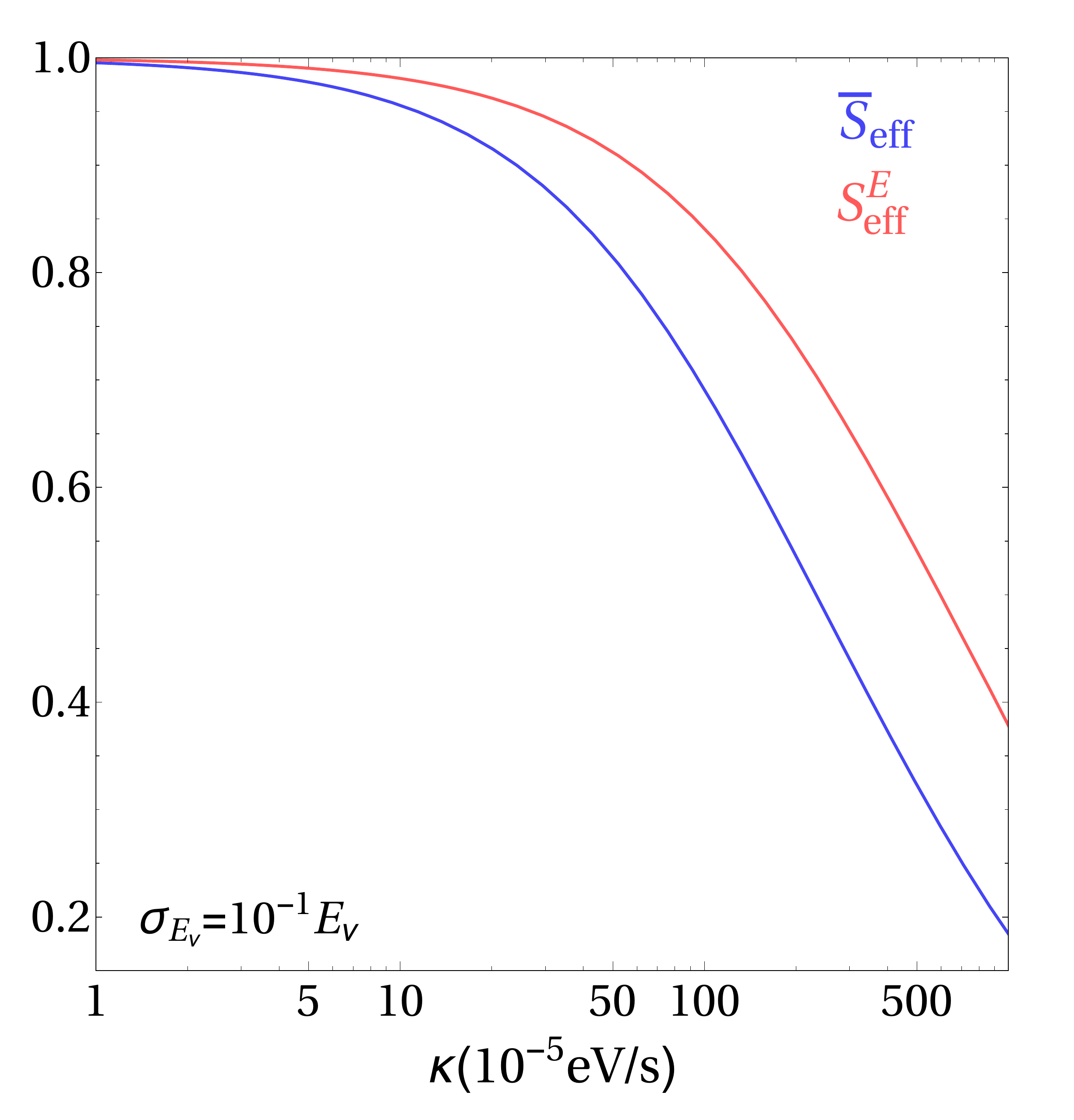}}
	\caption{Same as \cref{fig:1left,fig:BARSeffPD} for the case of 
	neutrino decay.}
	\label{fig:SeffDecay}
\end{figure}

 Effectively, when comparing \cref{eq:P_used} with \cref{eq:P_used_exp}, the transition probability in the invisible decay model is obtained by making a replacement $\cos^2 (\Delta m_i^2 /4...)\to \text{Exp}[-\kappa_i...]\equiv \text{Exp}[-\Phi_D]$. Hence, the treatment for obtaining 
suppression factors ($S_\text{eff}$,$\bar{S}_\text{eff}$) for the neutrino decay model is analogous to the one presented in \cref{sec:flavor_ratio} by employing the above replacement.
In \cref{fig:SeffDecay} we show the analogue of \cref{fig:1left} and \cref{fig:BARSeffPD} for the case of neutrino decays. Both panels show a comparison between a single known blazar and a number of unidentified blazars assumed to follow the $R(z)$ distribution (hence the blue lines in both panels are obtained by making an integral over redshift, see \cref{eq:Seff}). In \cref{fig:appleft} we fix the energy to $E_\nu=0.29$ PeV and in \cref{fig:appright} the energy integral is performed in the range $(E_\nu^\text{min}, E_\nu^\text{max})=(0.1,5)$ PeV. From both panels we observe that the relevant values of $\kappa_i$, for which the suppression factors are not trivial (i.e. equal to $1$) are $\kappa \gtrsim 10^{-4}$ eV/s. As in the pseudo-Dirac scenario, we observe an apparent difference between the suppression factors when comparing the cases with identified sources and the assumed distribution of unknown sources. This essentially confirms our conclusions from \cref{sec:flavor_ratio} that studying known sources may alter the predictions on the viable values of the parameters $\Delta m^2$ (pseudo-Dirac) and $\kappa$ (neutrino decay).

\bibliographystyle{JHEP}
\bibliography{refs}

\end{document}